\documentstyle[preprint,pra,aps,epsf]{revtex}

\begin{document}

\tightenlines

\title{\bf Electron recombination with multicharged ions via
chaotic many-electron states}

\author{V. V. Flambaum$^1$, A. A. Gribakina\cite{EAIFHE}, G. F. Gribakin$^2$,
and C. Harabati$^1$}

\address{$^1$School of Physics, The University of New South Wales, Sydney, 
UNSW 2052, Australia}

\address{$^2$Department of Applied Mathematics and Theoretical Physics, 
Queen's University, Belfast BT7 1NN, UK}

\date{\today }

\maketitle
\vspace{1cm}

\begin{abstract}
We show that a dense spectrum of chaotic multiply-excited 
eigenstates can play a major role in collision processes involving 
many-electron multicharged ions. A statistical theory based on
chaotic properties of the eigenstates enables one to obtain
relevant energy-averaged cross sections in terms of sums over
single-electron orbitals. Our calculation of the low-energy electron
recombination of Au$^{25+}$ shows that the resonant process is
200 times more intense than direct radiative recombination, which
explains the recent experimental results of Hoffknecht {\em et al.}
[J. Phys. B {\bf 31}, 2415 (1998)].
\end{abstract}
\vspace{1cm}


In this paper we give a quantitative explanation of the puzzle of
electron recombination with Au$^{25+}$. We also demonstrate how to calculate
the contribution of ``chaotic'' multiply-excited states of the compound ion,
which mediate electron recombination with complex many-electron ions.

Experimentally, this process was studied recently at the UNILAC heavy ion
accelerator facility of the GSI in Darmstadt \cite{Hoffknecht:98}.
In spite of a high energy resolution \cite{comment1} the measured
recombination rate did not reveal any resonances and only showed two broad
structures around 30 and 80 eV. However, its magnitude at low electron
energies $\varepsilon \sim 1$~eV exceeded the radiative recombination (RR)
rate by a factor of 150 \cite{comment2}, although the observed energy 
dependence was close to to that of RR.

It is well-known that recombination rate can be enhanced by dielectronic
recombination (DR). In this process the incident electron is
captured in a doubly-excited state of the compound ion, which is then
stabilised by photoemission. Suggested originally by J. Sayers, it was first
considered by Massey and Bates
\cite{Massey:43} in the problem of ionospheric oxygen.
Later DR was found to be important for the ionization balance in
the solar corona and high-temperature plasmas on the whole \cite{Burgess:64}.
Electron-ion recombination has been measured directly in the laboratory
since early 1980's \cite{Recomb:92}. More recently the use of heavy-ion
accelerators and electron coolers of ion storage rings has greatly advanced
the experiment \cite{Andersen:89,Kilgus:90}. Recombination rates for many
ions have been measured from threshold to hundreds of eV electron energies
with a fraction-of-eV resolution
\cite{Schennach:94,Gao:95,Schuch:96,Uwira:96,Uwira:97,Zong:97,Mannervik:98}.
For few-electron ions the measured rates were in good agreement with
theory which added the contribution of DR resonances to the RR background,
e.g., He$^+$ \cite{DeWitt:94}, Li-like C$^{4+}$ \cite{Mannervik:98}
and Ar$^{15+}$ \cite{Schennach:94,Zong:97}, and B-like Ar$^{13+}$
\cite{DeWitt:96}. However, more complicated ions, e.g.,
Au$^{51+}$ \cite{Uwira:97} and U$^{28+}$ \cite{Uwira:96},
showed complicated resonance spectra and strongly-enhanced recombination
rates at low electron energies. In particular, in U$^{28+}$ the theory was
able to explain the main resonant features in the range 80--180 eV, but
failed to identify the resonances and reproduce the rate
at smaller energies \cite{Mitnik:98}. The situation in
Au$^{25+}$ \cite{Hoffknecht:98} looks even more puzzling.

\begin{figure}[h]
\epsfxsize=13cm
\centering\leavevmode\epsfbox{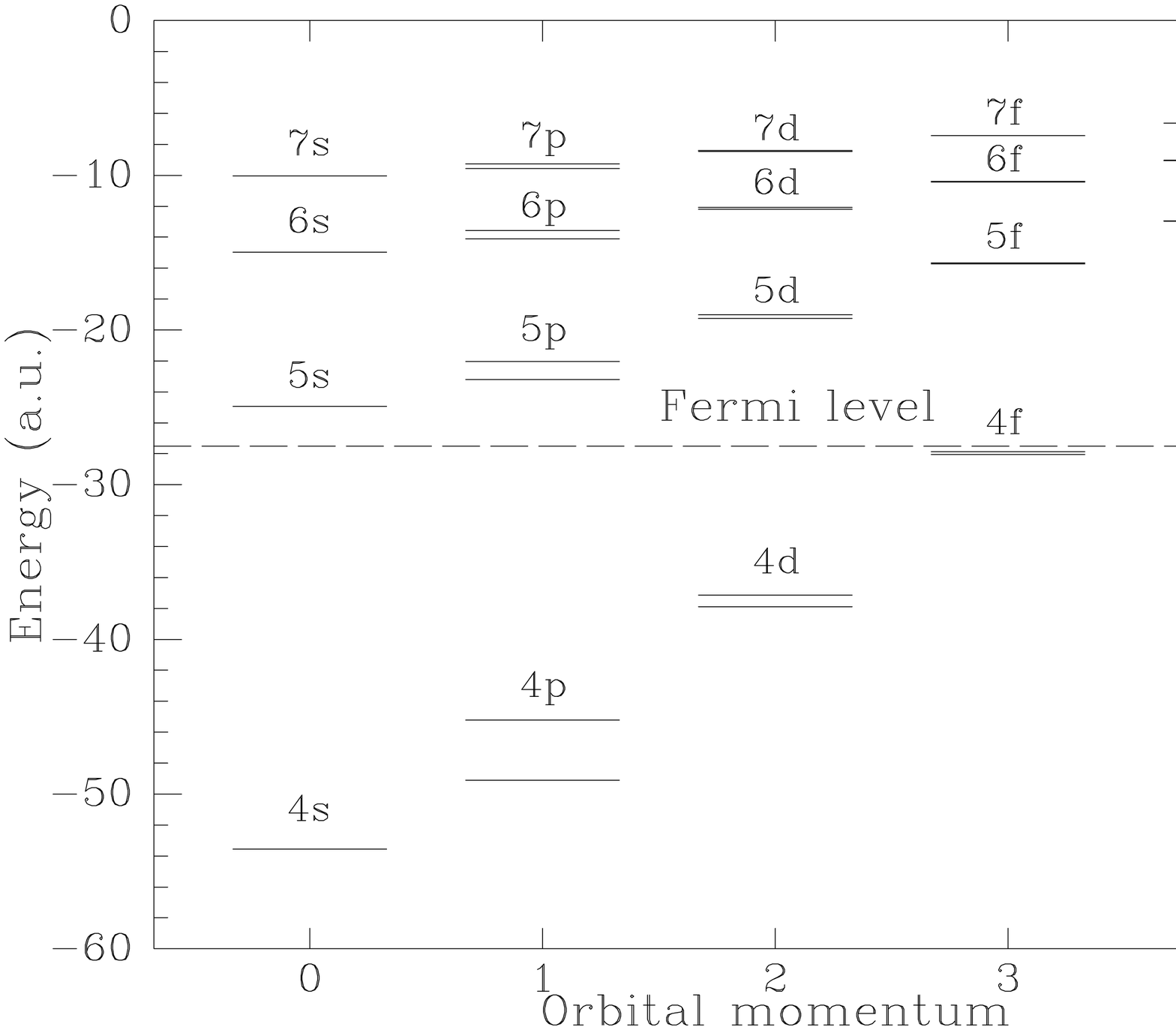}
\vspace{8pt}
\caption{Energies of the occupied and vacant single-particle
orbitals of Au$^{24+}$ obtained in the Dirac-Fock calculation.}
\label{fig:orb}
\end{figure}

In Ref. \cite{Au} we suggested that electron recombination with Au$^{25+}$
is mediated by complex {\em multiply-excited} states of Au$^{24+}$,
rather than ``simple'' dielectronic resonances. Electrons could be captured
in these states due to a strong configuration interaction in this open-shell
system (the ground state of Au$^{24+}$ is $4f^9$). The single-particle
spectrum of Au$^{24+}$ does not have large gaps, see Fig. \ref{fig:orb}.
Using the single-particle orbitals we generated many-electron
configurations, evaluated their energies and estimated the energy density of
multiply-excited states \cite{Au}. Owing to the ``gapless''
single-particle spectrum, the density increases rapidly as a function of
energy, as described by the Fermi-gas-model ansatz
\cite{Bohr:69}
\begin{equation}\label{eq:rho}
\rho (E)=AE^{-\nu }\exp (a \sqrt{E}),
\end{equation}
with $A=31.6$, $\nu =1.56$, $a=3.35$ a.u. \cite{comment3}, where $E$ is the
energy above the ground state in atomic units used throughout the paper.

The excited states are characterised by their total angular momentum and
parity $J^\pi $, and are $2J+1$ times degenerate. Therefore the total level
density is a sum of partial level densities:
$\rho (E)=\sum _{J^\pi }(2J+1)\rho _{J^\pi }(E)$. The spectrum of Au$^{24+}$
near the ionization threshold, $E=I\approx 27.5$ a.u., contains $J$
from $\frac{1}{2}$ to $\frac{35}{2}$ \cite{Au}. Their distribution is in
agreement with statistical theory, $\rho _{J^\pi }\propto f(J)$, where
\begin{equation}\label{eq:f_J}
f(J)= \frac{2(2J+1)}{(2J_m+1)^2}\exp \left[ -\frac{(2J+1)^2}
{2(2J_m+1)^2}\right] ,
\end{equation}
$J_m$ being the most abundant value of $J$ \cite{Bohr:69,Bauche:87}.
Numerically we find $J_m \approx \frac{9}{2}$. Using Eq. (\ref{eq:f_J}) we
can estimate the partial densities by $\rho _{J^\pi }=f(J)\rho /
\langle 2J+1\rangle $, where $\langle 2J+1\rangle $ is an average over 
$f(J)$. For the most abundant $J$ values this leads to
$\rho _{J^\pi }(E)=A_{J^\pi }E^{-\nu }\exp (a \sqrt{E})$
with $A_{J^\pi }\approx 0.15$. Near the ionization threshold we have
$\rho _{J^\pi }\approx 3.6\times 10^4$ au, which means that the
spacing between the multiply-excited states with a given $J^\pi $ is very
small: $D=1/\rho _{J^\pi }\sim 1$ meV. This would explain why autoionising
resonances could not be observed in electron recombination with
Au$^{25+}$ \cite{Au}. However, the large density is only a ``kinematic'' 
reason behind the experimental finding, because we have not proved that
the electron can actually be captured in the multiply excited states.
In what follows we analyse the dynamics of electron capture and
show that the Coulomb interaction between the electrons makes the capture
efficient and accounts for the observed enhanced recombination rate.

Taking into account this interaction is the key problem in many-electron
processes.
In general, this can be achieved by constructing a basis of many-electron
states $\Phi _k $ from single-particle orbitals, and solving
the eigenproblem for the Hamiltonian matrix
$H_{ik}=\langle \Phi _i|\hat H|\Phi _k\rangle $, which yields the
eigenvalues $E_\nu $ and eigenstates $|\Psi _\nu \rangle =\sum _kC^{(\nu )}_k
|\Phi _k \rangle $
of the system (configuration interaction method). For open-shell systems with
a few valence electrons, e.g., rare-earth atoms, this becomes an increasingly
difficult task. The
density of states grows very rapidly with the excitation energy, and finding
the eigenstates requires diagonalisation of ever greater matrices.

On the other hand, when the level density is high and the two-body interaction
is sufficiently strong the system is driven into a regime of {\em many-body
quantum chaos}, where the effect of configuration mixing can be described
statistically. In this case each eigenstate contains a large number $N$ of
{\em principal} components $C^{(\nu )}_k\sim 1/\sqrt{N}$, corresponding
to the basis states $\Phi _k $ which are strongly mixed together.
This strong mixing takes place in a certain energy range $|E_k-E_\nu |\lesssim
\Gamma _{\rm spr}$, where $E_k\equiv H_{kk}$ is the mean energy of the basis
state and $\Gamma _{\rm spr}$ is the so-called {\em spreading width}.
More precisely, the mean-squared value of $C^{(\nu )}_k$
as a function of $E_k-E_\nu $, is described by a Breit-Wigner formula
\begin{equation}\label{eq:C^2}
\overline {\left| C_k^{(\nu)}\right|^2}=N^{-1}\frac{\Gamma _{\rm spr}^2/4}
{(E_k-E_\nu )^2+\Gamma _{\rm spr}^2/4},
\end{equation}
with $N=\pi \Gamma _{\rm spr}/2D$ fixed by normalisation
$\sum _k\left| C_k^{(\nu)}\right|^2\simeq
\int \overline {\left| C_k^{(\nu)}\right|^2} dE_k/D=1$.
The decrease of $C^{(\nu )}_k$ for $|E_k-E_\nu |>\Gamma _{\rm spr}$ is a
manifestation of perturbation theory: the admixture of distant basis states is
suppressed by large energy denominators.
% This relation can also be interpreted by saying that due to the interaction
% between the particles the basis states $\Phi _k $ are nonstationary.
% They decay, or spread over the eigenstates, with a typical time
% $\tau =\hbar /\Gamma _{\rm spr}$.
Apart from this systematic variation the components $C^{(\nu )}_k$ behave as
Gaussian random variables.

This picture of many-body quantum chaos is supported by numerical studies of
nuclei \cite{Zel}, complex atoms \cite{Ce} and ions \cite{Au}.
In particular, Ref. \cite{Au} provides an estimate of the spreading width in
Au$^{24+}$: $\Gamma _{\rm spr}\approx 0.5$ a.u. Hence, a typical
eigenstate near the ionization threshold contains $N\sim 2\times 10^4$
principal components.

For low-energy electrons the contribution of multiply-excited autoionising
states (resonances) to the recombination cross section is \cite{Landau}
\begin{equation}\label{eq:sigma_res}
\sigma _r=\frac{\pi }{k^2}\sum _\nu \frac{2J+1}{2(2J_i+1)}\,
\frac{\Gamma _\nu ^{(r)}\Gamma _\nu ^{(a)}}{(\varepsilon -\varepsilon _\nu )^2
+\Gamma _\nu ^2/4},
\end{equation}
where $\varepsilon =k^2/2$ is the electron energy, $J_i$ is the angular
momentum of the initial (ground) target state, $J$ are the angular momenta
of the resonances, $\varepsilon _\nu =E_\nu -I$ is the position of the
$\nu $th resonance relative to the ionization threshold of the compound
(final-state) ion, and $\Gamma _\nu ^{(a)}$, $\Gamma _\nu ^{(r)}$,
and $\Gamma _\nu =\Gamma _\nu ^{(r)}+\Gamma _\nu ^{(a)}$ are its
autoionisation, radiative, and total widths, respectively
\cite{comment4}. When the resonance spectrum is dense, $\sigma _r$
can be averaged over an energy interval $\Delta \varepsilon$,
$D\ll \Delta \varepsilon \ll \varepsilon $, yielding
\begin{equation}\label{eq:sigres_av}
\bar \sigma _r=\frac{2\pi ^2}{k^2}\sum _{J^\pi }
\frac{2J +1}{2(2J_i+1)D}\left\langle \frac{\Gamma _\nu ^{(r)}
\Gamma _\nu ^{(a)}}{\Gamma _\nu ^{(r)}+\Gamma _\nu ^{(a)}} \right\rangle ,
\end{equation}
where $\langle \dots \rangle $ means averaging. The fluorescence yield
$\omega _f\equiv \Gamma _\nu ^{(r)}/(\Gamma _\nu ^{(r)}+\Gamma _\nu ^{(a)})$,
fluctuates weakly from resonance to resonance (see below), which allows one to
write $\bar \sigma _r=\bar \sigma _c\omega _f$, where
\begin{equation}\label{eq:sig_cap}
\bar \sigma _c=\frac{\pi ^2}{k^2}\sum _{J^\pi } \frac{(2J +1)
\Gamma ^{(a)}}{(2J_i+1)D}
\end{equation}
is the energy-averaged capture cross section, and $\Gamma ^{(a)}$ is the
average autoionisation width.

Unlike complex multiply-excited states $\Psi _\nu $, the initial state of
the recombination process is simple. It describes an electron with the
energy $\varepsilon $ incident on the ground state $\Phi _i$ of the target
(e.g., Au$^{25+}\,4f^8$, $J_i=6$). The autoionisation width is then given
by perturbation theory as
\begin{eqnarray}\label{eq:Gamma_a}
\Gamma _\nu ^{(a)}&=&2\pi |\langle \Psi _\nu |\hat V|
\Phi _i;\varepsilon \rangle |^2 \\
&=&2\pi \sum _{k,k'} C_k^{(\nu)*}C_{k'}^{(\nu)}
\langle \Phi _i;\varepsilon | \hat V|\Phi _{k'}\rangle
\langle \Phi _k|\hat V|\Phi _i;\varepsilon \rangle ,\nonumber
\end{eqnarray}
where $\hat V$ is the electron Coulomb interaction, and the continuum
states $\varepsilon $ are normalised to unit energy interval. Averaging
$\Gamma _\nu ^{(a)}$ over the chaotic states $\nu $ with
$E_\nu \approx I+\varepsilon $, we
make use of the fact that their components are random and uncorrelated, which
leads to
\begin{equation}\label{eq:Gamma_a1}
\Gamma ^{(a)}=2\pi \sum _k \overline {\left| C_k^{(\nu)}\right|^2}
|\langle \Phi _k |\hat V|\Phi _i;\varepsilon \rangle |^2 .
\end{equation}
Being a two-body operator, $\hat V$ can move only two electrons at a time.
A nonzero contribution to $\Gamma ^{(a)}$ is given by the basis states which
differ from the initial state $|\Phi _i; \varepsilon \rangle $ by the
positions of two electrons. Therefore, in Eq. (\ref{eq:Gamma_a1}) we only
need to sum only over {\em doubly-excited} basis states $\Phi _d$. With the
help of Eq. (\ref{eq:C^2}) the capture cross section (\ref{eq:sig_cap}) becomes
\begin{equation}\label{eq:sig_cap1}
\bar \sigma _c=\frac{\pi }{k^2}\sum _d \frac{2J +1}
{2(2J_i+1)}\frac{\Gamma _{\rm spr}2\pi 
|\langle \Phi _d |\hat V|\Phi _i;\varepsilon \rangle |^2}
{(E_d-I-\varepsilon )^2+\Gamma _{\rm spr}^2/4}.
\end{equation}
This form makes it clear [cf. Eq. (\ref{eq:sigma_res})] that the two-electron
excitations $\Phi _d$ play the role of {\em doorway states} for the
electron capture process. Since these states are not the eigenstates of the
system they have a finite energy width $\Gamma _{\rm spr}$.

The wave function of a doorway state can be constructed using the
creation-annihilation operators, $|\Phi _d\rangle =a_\alpha ^\dagger
a_\beta ^\dagger a_\gamma |\Phi _i\rangle $, where $\alpha
\equiv n_\alpha l_\alpha j_\alpha  m_\alpha $ and
$\beta \equiv n_\beta l_\beta j_\beta m_\beta $ are
excited single-electron states, and $\gamma \equiv n_\gamma 
l_\gamma j_\gamma m_\gamma $  corresponds to a hole in the
target ground state. (Note that we are using relativistic Dirac-Fock orbitals
$nljm$.) Of course, to form doorway states with a given $J$ the
angular momenta of the electrons and ionic residue must be coupled into
the total angular momentum $J$. However, the $2J+1$ factor and summation
over $J$ implied in Eq. (\ref{eq:sig_cap1}) account for all possible
couplings, and we can simply sum over the single-electron excited states
$\alpha ,~\beta $ and hole states $\gamma $, as well as the partial waves
$lj$ of the continuous-spectrum electron $\varepsilon $. As a result, we have
\begin{eqnarray}\label{eq:sig_cap2}
\bar \sigma _c&=&\frac{\pi ^2}{k^2}\sum _{\alpha \beta \gamma ,lj}
\frac{\Gamma _{\rm spr}}{(\varepsilon -\varepsilon _\alpha -
\varepsilon _\beta +\varepsilon _\gamma )^2 +\Gamma _{\rm spr}^2/4}
\sum _\lambda \frac{\langle \alpha ,\beta \| V_\lambda \| \gamma ,
\varepsilon lj \rangle }{2\lambda +1}\nonumber \\
&\times &\Biggl[ \langle \alpha ,\beta \| V _\lambda \| \gamma ,
\varepsilon lj\rangle -(2\lambda +1) \sum _{\lambda '}
(-1)^{\lambda +\lambda '+1}\left\{ {\lambda \atop \lambda '}{j_\alpha 
\atop j_\beta }{j \atop j_\gamma }\right\} \langle \alpha ,\beta \| V
_{\lambda '}\| \varepsilon lj,\gamma \rangle \Biggr] ,
\end{eqnarray}
where $\varepsilon _\alpha $, $\varepsilon _\beta $ and $\varepsilon _\gamma $
are the orbital energies, and $\langle \alpha ,\beta \| V_\lambda \| \gamma ,
\varepsilon lj \rangle $ is the reduced Coulomb matrix element \cite{details}.
The above equation is directly applicable to targets with closed-shell ground
states. If the target ground state contains partially occupied orbitals,
a factor
\begin{equation}\label{eq:occup}
\frac{n_\gamma }{2j_\gamma +1}\left(1-\frac{n_\alpha }{2j_\alpha 
+1}\right)\left(1-\frac{n_\beta }{2j_\beta  +1}\right),
\end{equation}
where $n_\alpha $, $n_\beta $, and $n_\gamma $ are the orbital occupation
numbers in the ground state $\Phi _i$, must be introduced on the right-hand
side of Eq. (\ref{eq:sig_cap2}). Steps similar to those that lead to
Eq. (\ref{eq:sig_cap2}) were used to obtain mean-squared matrix elements of
operators between chaotic many-body states \cite{Ce,Flambaum:93}.

Chaotic nature of the multiply-excited states $\Psi _\nu $ can also be
employed to estimate their radiative widths $\Gamma _\nu ^{(r)}$. 
Electron-photon interaction is described by a single-particle dipole operator
$\hat d$. Any excited electron in $\Psi _\nu $ may emit a photon, thus leading
to radiative stabilisation of this state. The total photoemission rate
$\Gamma _\nu ^{(r)}$ can be estimated as a weighted sum of single-particle
rates,
\begin{equation}\label{eq:Gamma_r}
\Gamma _\nu ^{(r)}\simeq \sum _{\alpha ,\beta }
\frac{4\omega _{\beta \alpha }^3} {3c^3}
|\langle \alpha \|\hat d\|\beta \rangle |^2
\left\langle \frac{n_\beta }{2j_\beta +1}\left( 1-\frac{n_\alpha }
{2j_\alpha +1}\right) \right\rangle _\nu ,
\end{equation}
where $\omega _{\beta \alpha }=\varepsilon _\beta -\varepsilon _\alpha >0$,
$\langle \alpha \|\hat d\|\beta \rangle $ is the reduced dipole operator
between the orbitals $\alpha $ and $\beta $, and
$\langle \dots \rangle _\nu $ is the
mean occupation number factor. Since $\Psi _\nu $ have large numbers
of principal components $N$, their radiative widths display small $1/\sqrt{N}$
fluctuations. This can also be seen if one recalls that a chaotic
multiply-excited state is coupled by photoemission to many lower-lying states,
and the total radiative width is the sum of a large number of (strongly
fluctuating) partial widths. A similar effect is known in compound nucleus
resonances in low-energy neutron scattering \cite{Bohr:69}.

It is important to compare the radiative and autoionisation widths of
chaotic multiply-excited states. Equation (\ref{eq:Gamma_r}) shows that
$\Gamma ^{(r)}$ is comparable to the single-particle radiative widths.
On the other hand, the autoionisation width $\Gamma ^{(r)}$,
Eq. (\ref{eq:Gamma_a1}), is suppressed by a factor
$\left|C_k^{(\nu )}\right|^2\sim N^{-1}$ relative to that of a typical
dielectronic resonance. A comparison of
Eqs. (\ref{eq:sig_cap}) and (\ref{eq:sig_cap1}) also shows that
$\Gamma ^{(a)}$ is suppressed as $D/\Gamma _{\rm spr}$. Therefore, in systems
with dense spectra of chaotic multiply-excited states the autoionisation
widths are small. Physically this happens because the coupling strength
of a two-electron doorway state is shared between many complex eigenstates.
As a result, the radiative width may dominate in the total width of the
resonances, $\Gamma ^{(r)}\gg \Gamma ^{(a)}$, making their
fluorescence yield close to unity. Our numerical results for the
recombination of Au$^{25+}$ presented below, confirm this scenario.

The resonant recombination cross section should be compared to the 
direct radiative recombination cross section
\begin{equation}\label{eq:sigmad}
\sigma _d= \frac{32\pi }{3\sqrt{3}c^3}\,\frac{Z_i^2}
{k^2} \ln \left( \frac{Z_i}{n_0k}\right) ,
\end{equation}
obtained from the Kramers formula by summing over the principal
quantum number of the final state \cite{Au}. Here $Z_i$ is the ionic charge
($Z_i=25$ for Au$^{25+}$), and $n_0$ is the principal quantum number of
the lowest unoccupied ionic orbital. Note that the direct 
and resonant recombination cross sections of Eqs. (\ref{eq:sigmad})
and  (\ref{eq:sigres_av}) have similar energy dependences. Therefore, 
for the purpose of comparing with experiment we can evaluate the cross 
sections at one low energy, say $\varepsilon =0.5$ eV. This energy is 
much greater than the thermal energy spread of the electron beam
\cite{comment1} and the cross sections can be compared directly to the
experiment \cite{Hoffknecht:98}.

Using Eqs. (\ref{eq:sig_cap2}) and (\ref{eq:occup}) and summing over the
orbitals in Fig. \ref{fig:orb} and electron partial waves up to $h_{11/2}$,
we obtain the capture cross section $\sigma _c=23\times 10^{-16}$~cm$^2$
at $\varepsilon =0.5$ eV. A comparison with Eq. (\ref{eq:sig_cap}) 
shows that the sum which contains the autoionisation width, is
\begin{equation}\label{eq:GammaD}
\sum _{J^\pi } \frac{(2J +1)
\Gamma ^{(a)}}{(2J_i+1)D}=0.305.
\end{equation}
Combining this with $D=3\times 10^{-5}$ a.u. and taking into account that
about ten different $J$ and two parities contribute to the sum, we obtain
$\Gamma ^{(a)}\sim 5\times 10^{-7}$ a.u. On the other hand, a numerical
calculation of Eq. (\ref{eq:Gamma_r}) gives
$\Gamma ^{(r)}=3\times 10 ^{-5}$ a.u. Therefore, $\Gamma ^{(r)}\gg
\Gamma ^{(a)}$ and $\omega _f\approx 1$. Hence, the resonant recombination
cross section is basically equal to the capture cross section:
$\sigma _r\approx 23\times 10^{-16}$~cm$^2$. This value is in
good agreement with the experimental $\sigma ^{\rm (exp)}=27\times 
10^{-16}$~cm$^2$ \cite{Hoffknecht:98}, and exceeds the direct recombination
cross section (\ref{eq:sigmad}), $\sigma _d=0.12\times 
10^{-16}$~cm$^2$, by a factor of 200.

In summary, we have shown that a dense spectrum of chaotic multiply-excited
states can play a major role in the dynamics of electron 
recombination with many-electron multicharged ions, and possibly
other processes, e.g. charge transfer in collisions of multiply-charged
ions with neutral atoms. Based on the chaotic nature of these states, we have
developed a statistical theory which enables one to calculate energy-averaged
cross sections for processes which go via such resonances. Applied to the
recombination of Au$^{25+}$, the theory shows that the contribution of
resonances exceeds that of direct radiative recombination 200 times,
which explains the recent experimental findings \cite{Hoffknecht:98}.

We thank Prof. A. M\"uller for providing experimental data in numerical form.

%****************************************************************************

\figure

\begin{thebibliography}{10}
\bibitem{comment1} The experimental energy resolution in
Ref. \cite{Hoffknecht:98} was limited by the transversal temperature of the
electron beam $kT_\perp =0.1$~eV.
\bibitem{comment3}This fit is valid for $E>1$ a.u. In the Fermi-gas model
$a$ is related to the single-particle level density at the Fermi level
$g(\varepsilon _F)=3a^2/2\pi^2$. The fitted value $a=3.35$ gives
$g(\varepsilon _F)=1.7$ a.u., in agreement with the orbital spectrum in
Fig. \ref{fig:orb}.
\bibitem{comment4} Here we assume that the electron energy is below the target
excitation threshold.
\bibitem{details}A detailed derivation will be published elsewhere.
\end{thebibliography}
\end{document}